\documentstyle[12pt]{article}

\def\s2m{$OSp(1/2m,R)$}
\def\s32{$OSp(1/32,R)$}
\def\s64{$OSp(1/64,R)$}
\def\2m{$Sp(2m,R)$}
\def\32{$Sp(32,R)$}
\def\64{$Sp(64,R)$}

\def\s32l{$OSp(1/32,R)_L$}
\def\s32r{$OSp(1/32,R)_R$}
\def\s32lr{$OSp(1/32,R)_{L-R}$}

\newcommand{\eq}{\begin{equation}}
\newcommand{\en}{\end{equation}}
\newcommand{\eqn}{\begin{eqnarray}}
\newcommand{\enn}{\end{eqnarray}}

\begin{document}
\begin{center}
{\bf Generalized AdS/CFT Dualities and Space-Time Symmetries of
M/Superstring Theory}
\\
\vskip .5cm

{\bf  Murat G\"{u}naydin}\footnote{Work supported in part by the
National Science Foundation under Grant Number PHY-9802510. \newline
e-mail: murat@phys.psu.edu} \\
\vskip .5cm
 Penn State University \\
 Physics Department\\
University Park, PA 16802

\vskip 1cm

{\bf Abstract}

\end{center} 

I review the relationship  between AdS/CFT ( anti-de Sitter /
 conformal field theory)
 dualities and the general theory of unitary lowest weight
  (ULWR) (positive energy) representations of
non-compact space-time groups and supergroups. The ULWR's have the
 remarkable property that they can be constructed by tensoring some
 fundamental ULWR's ( singletons or doubletons).  Furthermore, one can
go from the manifestly unitary compact basis of the ULWR's
 of the conformal group ( Wigner picture) to the manifestly covariant
 coherent state basis ( Dirac picture) labelled by the space-time
  coordinates.  Hence every irreducible ULWR corresponds to a covariant
 field with a  definite conformal dimension.
 These results extend to higher dimensional  generalized
 spacetimes ( superspaces) defined by Jordan (super) algebras and Jordan
(super)  triple systems. In particular, they extend to the ULWR's of the
M-theory symmetry superalgebra $OSp(1/32,\mathbf{R})$.

\vskip 2cm

{\it Invited talk given at the IXth Marcel Grossmann Meeting in Rome,
July, 2000}

\newpage

\section{Introduction}
The conjecture of Maldacena \cite{mald,ew98,gkp,agmoo}  on the duality between
 the large $\cal N$ limits
of certain conformal field theories (CFT) in $d$ dimensions and the
 superstring theory on the products of $d+1$ dimensional
anti-de Sitter (AdS) spaces with spheres fits perfectly  into the
general theory of the construction of positive energy representations of
AdS and conformal groups and supergroups by a simple  tensoring procedure from
some "fundamental" representations as reviewed in \cite{mgdm,mg98}.
These "fundamental" representations have
been called singletons or doubletons and act like the basic building
blocks ( quarks) of all the positive energy representations. For
simple noncompact groups including AdS and conformal groups the general theory was
developed in \cite{gs}. The extension of the general theory to noncompact
supergroups including AdS and conformal supergroups was given in
\cite{bg}.

The supermultiplets of fields of the superconformal field theories to which the superstring/M-theory is
dual to in various dimensions are precisely  the "CPT self-conjugate"
singleton or doubleton supermultiplets of the corresponding space-time
supergroups.

This simple tensoring procedure was applied to the construction of the
Kaluza-Klein spectra of ten and eleven dimensional supergravity theories
long time ago.
In \cite{gw}  the spectrum of the $S^{7}$ compactification
of  eleven dimensional supergravity was shown to fit into an infinite tower of short
unitary supermultiplets of $OSp(8/4,R)$. The ultra-short singleton supermultiplet of
$OSp(8/4,R)$ sits at the bottom of this infinite tower of Kaluza-Klein modes and
 decouple from the spectrum as local gauge degrees of freedom
\cite{gw}.
However , even though it
decouples from the spectrum as local gauge modes, one can
generate the entire spectrum of 11-dimensional supergravity over $S^7$
by tensoring $p$ copies (``colors'') ($p=2,3,4,\ldots $)  of  singleton
 supermultiplets  and restricting oneself to "CPT self-conjugate "
supermultiplets.

The spectrum of 11-d supergravity over the four
sphere $S^{4}$ \cite{gnw,ptn}  was shown to  fit into
an infinite tower of unitary supermultiplets
of $OSp(8^{*}/4)$ with the even subgroup $SO(6,2)\times USp(4)$ in \cite{gnw}.
Again the vacuum doubleton supermultiplet of $OSp(8^{*}/4)$ decouples from the
spectrum as local gauge degrees of freedom
 and the entire physical spectrum of 11-dimensional
supergravity over $S^4$ was obtained by simply tensoring an arbitrary number  (colors)
of these doubleton supermultiplets  and restricting oneself to the vacuum
("CPT self-conjugate )
supermultiplets \cite{gnw}.

The spectrum of the $S^{5}$ compactification of ten dimensional
IIB supergravity was calculated in \cite{gm,krv}.
The entire spectrum
falls into an infinite tower of massless and massive unitary supermultiplets of
$N=8$ $AdS_{5}$ superalgebra $SU(2,2/4)$ \cite{gm}.
The "CPT self-conjugate" doubleton supermultiplet of
$N=8$ $AdS$ superalgebra  decouples from the physical
spectrum as local gauge degrees of freedom.
By tensoring it with itself
repeatedly and restricting oneself to the $CPT$ self-conjugate vacuum
supermultiplets one generates the entire spectrum of
Kaluza-Klein states of ten dimensional IIB supergravity on $S^{5}$.

As was pointed out in \cite{gm,mgnm2}  the CPT self-conjugate $N=8$  $AdS_5$
doubleton supermultiplet does not have a Poincar\'{e} limit in five dimensions
and its field theory is the conformally invariant $N=4$ super Yang-Mills theory in
$d=4$ Minkowski space which can be identified as the boundary of $AdS_5$.

Similarly, the singleton supermultiplet of $OSp(8/4,R)$ and the doubleton
supermultiplet of $OSp(8^*/4)$ do not have a Poincar\'{e} limit in $d=4$ and
$d=7$, respectively, and their field theories are conformally invariant theories
in one lower dimension \footnote{ see \cite{mgdm} for references }.
Thus we see that at the level of physical states the proposal of Maldacena
is perfectly consistent with  the above
mentioned results.
\section{ Massless and Massive Supermultiplets of Anti-de Sitter Supergroups}
The Poincar\'{e} limit of the remarkable representations (singletons)
 of the $d=4\  AdS$ group
$SO(3,2)$  discovered by Dirac \cite{pad} are known to be singular \cite{cf1}.
 However, the tensor product of
 two singleton representations decomposes into an infinite set of
massless unitary  irreducible
 representations which do have a
smooth Poincar\'{e} limit \cite{cf1,gs,mg81}.
 Similarly, the tensor product of two
singleton supermultiplets of $N$ extended $AdS_4$ supergroup $OSp(N/4,R)$
decomposes into an infinite set of massless supermultiplets which do have a Poincar\'{e}
limit in $AdS_5$ \cite{mg81,gw,gh,mg89}.
In contrast to $AdS_4$ group $SO(3,2)$, the $AdS_5$ group $SU(2,2)$ and
$AdS_7$ group $SO(6,2)$ admit infinitely many "remarkable" representations
that do not have a Poincar\'{e} limit in 5 and 7 dimensions , respectively.
These representations have been called " doubletons" since they require
two sets of oscillators for their realization \cite{gnw,gm}.
 The doubleton supermultiplets of extended $AdS$ supergroups in $d=5\
 (SU(2,2/N))$ and $d=7\ (OSp(8^*/2N))$ share the same remarkable features of the singleton
supermultiplets of $d=4$ $AdS$ supergroups i.e  the tensor product of any two
doubletons decompose into an infinite set of massless supermultiplets
\cite{gnw,gm,mg89,gmz1,gmz2,mgst}.
 In $d=3$ the $AdS$ group $SO(2,2)$ is not simple and is isomorphic to
$SO(2,1)\times SO(2,1)$. Each $SO(2,1)$ factor can be extended to a simple
superalgebra with some internal symmetry group and  one has a rich variety of $AdS$
supergroups in $d=3$ whose unitary supermultiplets were studied in\cite{gst}.

The tensor product of more than two copies of the singleton or doubleton
supermultiplets of $AdS$ supergroups decompose into an infinite set of {\it
massive}
supermultiplets
\cite{gnw,gw,gm,mg98,gmz1,gmz2,mgst,mg90,mgrs,mg88,cgkrz} . The Kaluza-Klein spectrum of the
11-dimensional and ten dimensional IIB supergravity compactified over $S^7(S^4)$
and $S^5$ fall into CPT self-conjugate short massless and massive supermultiplets
of $OSp(8|4,R) (OSp(8^*|4))$ and $SU(2,2|4)$, respectively. As
explained above they can all be obtained by tensoring  $p$
copies ($p=1,2,3,...$) of
 the corresponding singleton or doubleton supermultiplets and restricting
 oneself to CPT-self conjugate irreducible multiplets. If one does not
 restrict oneself to self-conjugate supermultiplets then tensoring
 procedure yields intermediate as well as long massive supermultiplets of
 the corresponding supergroups. These intermediate and long supermultiplets
are expected to descend  from the massive modes of M/Superstring theory
\cite{gmz1,gmz2,cgkrz}.
\section{Generalized Space-times}
\subsection{ Rotation, Lorentz and Conformal Groups of Generalized Spacetimes}
In the twistor formalism a coordinate vector  $x_{\mu}$ in  four-dimensional
 space-time   is naturally  represented
as a  $2\times 2$  Hermitian matrix $x=x_{\mu}\sigma^{\mu}$ over the field of complex
numbers ${\mathbf C}$ .
Since the Hermitian matrices over the field of complex
numbers close under the symmetric anti-commutator product
one  can thus regard the coordinate vectors as elements of a Jordan algebra
denoted as $J_2^{\mathbf C}$
\cite{mg75,mg80}.
  Then the rotation, Lorentz and conformal groups in
$d=4$ can be identified with the automorphism , reduced  structure
and M\"{o}bius ( linear fractional) groups of  the Jordan algebra of
  $J_2^{\mathbf C}$ \cite{mg75,mg80}.
Furthermore, this
interpretation allows one to define generalized space-times whose
coordinates are parametrized by the elements of  Jordan algebras
(or Jordan triple systems) \cite{mg75,mg80,mg92} .
The  rotation $Rot(J)$,
Lorentz $Lor(J)$ and conformal $Con(J)$ groups of these generalized
 space-times are then identified with the automorphism $Aut(J)$,
 reduced structure $Str_0(J)$
and M\"{o}bius  M\"{o}(J) groups of the corresponding Jordan
 algebra \cite{mg75,mg80,mg91,mg92}.
A complete list of the generalized spacetimes defined by formally real
Jordan algebras and triple systems (JTS)  as well as  their symmetry
groups
were given in \cite{mg91,mg92}.
\subsection{Conformal Fields over Generalized Spacetimes and
the Positive Energy Unitary Representations of
 Conformal Groups}
Remarkably,  the list of generalized conformal groups defined by
 simple formally real Jordan algebras and JTS's corresponds precisely to the list of
 simple non-compact groups that admit positive energy unitary
 representations (i.e unitary representations of the lowest weight
 type (ULWR)).
Furthermore,  the maximal compact
subgroups of the generalized conformal groups  are
simply the compact forms of their structure groups
(generalized Lorentz group times dilatations).
 For example,  the conformal group of the Jordan algebra
$J_2^{\mathbf C}$ corresponding to the four dimensional Minkowski space
 is $SU(2,2)$  with a maximal compact subgroup
$SU(2)\times SU(2)\times U(1)$ which is simply the compact form of the
structure group $SL(2,\mathbf C)\times SO(1,1)$.
In \cite{gmz2} it was
explicitly shown how to go from the  compact
$SU(2)\times SU(2) \times U(1)$ basis of the ULWR's of $SU(2,2)$ to the
manifestly covariant  $SL(2,\mathbf C)\times SO(1,1)$ basis. The transition from
the compact  to the covariant basis corresponds simply to going
from a "particle" basis to a coherent state basis of the corresponding
positive energy representation .
The coherent states are labelled by the elements of  $J_2^{\mathbf C}$
representing the coordinates of four dimensional Minkowski space. One can then
establish a one-to-one correspondence between irreducible ULWR's of
$SU(2,2)$ and the conformal fields transforming irreducibly under the Lorentz
group $SL(2,{\mathbf C}) $ with a definite conformal dimension \cite{gmz2,mack}.
Similarly, the conformal group $SO^*(8)$  of the Jordan algebra $J_2^{\mathbf H}$
parametrizing the six dimensional Minkowski space has a maximal compact subgroup
$U(4)$  which is the compact form of the
structure group $SU^*(4)\times SO(1,1)$. In \cite{mgst} it was shown
how to go from the  compact $U(4)$ basis of the ULWR's of $SO^*(8)$ to
the non-compact basis $SU^*(4) \times SO(1,1)$ which is simply the
Lorentz group in six dimensions times dilatations.
 The coherent states of the non-compact
basis are again labelled by the elements of the Jordan algebra $J_2^{\mathbf H}$
of $2\times 2$ hermitian quaternionic matrices representing the
coordinates of 6d Minkowski space.
Thus each irreducible
ULWR of $SO^*(8)$ can be identified with a field transforming
covariantly under the six dimensional Lorentz group $SU^*(4)$ with a definite conformal
dimension.
These results obtained explicitly for the conformal groups of $J_2^{\mathbf C}$
and $J_2^{\mathbf H}$ extend to the conformal groups of all formally
real Jordan algebras and of Hermitian Jordan triple systems \cite{mg2000,mg00}.
Recently, the nonlinear action of these conformal groups  have been generalized
to novel quasiconformal group actions over
certain natural extensions of the corresponding spacetimes \cite{gkn}.
\subsection{Generalized superspaces defined by Jordan superalgebras
and the ULWR's of their symmetry supergroups}
The generalized space-times defined by Jordan algebras extend naturally
to generalized superspaces over Jordan superalgebras
and super Jordan triple systems \cite{mg80,mg91,mg2000}.
One defines the  generalized superspaces by
multiplying the even elements of a Jordan superalgebra $J$
by real coordinates and their odd elements by Grassmann coordinates
\cite{mg80,mg91}.
The  rotation, Lorentz and
conformal supergroups of these generalized superspaces are then
 given by  the automorphism, reduced structure
and M\"{o}bius supergroups of $J$. A complete list of these supergroups
was given in \cite{mg91}.
The conformal groups of formally real Jordan algebras all admit ULWR's
and as explained above one can associate with each irreducible ULWR  a
 covariant conformal field with a definite conformal dimension.
 The corresponding real forms of  Jordan superalgebras or super Jordan triple systems
are such that their conformal supergroups admit positive energy
unitary representations. The general theory for the construction
of  the unitary lowest weight representations of non-compact supergroups was given in
\cite{bg} , both in a compact particle state basis as well as the compact
super-coherent state basis. The coherent states defined in
\cite{bg} for non-compact groups $G$ are
labelled by the complex variables parametrizing the hermitian symmetric
space $G/H$ where $H$ is the maximal compact subgroup. On the other hand
the coherent states defined in \cite{gmz2} for $SU(2,2)$ and in
\cite{mgst,fgt} for $OSp(8^*|4)$ as well as their generalizations to all
non-compact groups discussed in the previous section are labelled by
{\it real} (generalized) coordinates of the (generalized) space-times on which
$G$ acts as a (generalized) conformal group.

The irreducible ULWR's of (generalized) superconformal groups  correspond simply to a
supermultiplet of fields transforming irreducibly under the
(generalized) Lorentz group with definite conformal dimensions.
In a non-compact supercoherent state basis these unitary
supermultiplets are represented by superfields \cite{fgt} \footnote{ Recently,
a number of  papers studied the unitary
supermultiplets of conformal supergroups in 3,4 and 6 dimensions
using the formalism of  superfields
\cite{superfields}.}. The labels of supercoherent states are the
coordinates
of corresponding (generalized) superspaces.
\section{Unitary Supermultiplets of M-theory Superalgebra
$OSp(1|32,R)$}
The simple supergroup $OSp(1/32,R)$
 was proposed as the generalized $AdS$ supergroup in $d=11$ long time ago \cite{toine}.
One can also regard  $OSp(1/32,R)$ as the generalized conformal group
 in ten dimension.  $Sp(32,R)$ is  the conformal
 group of the Jordan algebra $J_{16}^{\mathbf{R}}$ of $16\times 16$ symmetric real
 matrices. As such one can consider  $OSp(1/32,R)$ as the
  conformal supergroup of a generalized superspace of  the type we
 discussed above. The unitary supermultiplets of  $OSp(1/32,R)$ were
 studied in \cite{mg98} with the aim of extending the $AdS/CFT$
 duality to the maximal possible dimension in M-theory.
The parity invariance of M-theory \cite{hw} requires the
 extension of $OSp(1/32,R)$ to a larger supergroup that admits parity invariant
  representations. The "minimal" such parity
symmetric supergroup is $OSp(1/32,R)_L\times OSp(1/32,R)_R$  \cite{horava}.
The two factors of $OSp(1/32,R)_L \times OSp(1/32,R)_R$ correspond to the embedding of
left-handed and right-handed spinor representations of $SO(10,2)$ in the fundamental
 representation of $Sp(32,R)$.
The contraction to the Poincar\'{e} superalgebra with central charges proceeds via a diagonal
subsupergroup $OSp(1/32,R)_{L-R}$ which contains the common subgroup $SO(10,1)$
of the two $SO(10,2)$ subgroups. The parity invariant tensor product of the singleton
 supermultiplets of the two factors decomposes into an infinite set of "doubleton"
 supermultiplets of the diagonal $OSp(1/32,R)_{L-R}$. There is a unique
 "CPT self-conjugate" doubleton supermultiplet whose tensor product with itself leads
 to "massless" supermultiplets. The "CPT self-conjugate" massless graviton supermultiplet
  contains fields corresponding to those of 11-dimensional supergravity plus
 additional ones.
 One then expects the corresponding   doubleton field theory to be a
 generalized  superconformal
 field theory in ten dimensions that is dual to an AdS phase of M-theory in the same
  sense as the duality between the $N=4$ super Yang-Mills in $d=4$ and the $IIB$
  superstring over $AdS_5 \times S^5$.     By tensoring more than two
 copies of the doubleton supermultiplet one obtains the massive
 supermultiplets of the conjectured AdS phase of M-theory.


\end{document}